\documentclass[10pt,conference]{IEEEtran}

\AtBeginDocument{%
  \providecommand\BibTeX{{%
    \normalfont B\kern-0.5em{\scshape i\kern-0.25em b}\kern-0.8em\TeX}}}

\usepackage[T1]{fontenc}
\usepackage{algorithmic}
\usepackage{amssymb}
\usepackage{amsmath}
\usepackage{balance}
\usepackage{graphicx}
\usepackage{textcomp}
\usepackage{listings}
\usepackage{url}
\usepackage{todonotes}
\usepackage[hidelinks]{hyperref}
\hypersetup{bookmarksdepth=-2}

\usepackage[super]{nth}
\usepackage{microtype}
\usepackage{cite}

\usepackage{booktabs}
\usepackage{csquotes}

\usepackage{acronym}
\acrodef{API}{application programming interface}
\acrodef{APR}{automatic program repair}

\usepackage{color}
\usepackage{xcolor}

\usepackage{xspace}

\definecolor{ckeywordcolor}{RGB}{127,0,85}
\definecolor{cstringcolor}{RGB}{42,0,255}
\definecolor{ccommentcolor}{RGB}{63,127,95}

\newcommand{\ovgu}{Otto-von-Guericke University\\ Magdeburg}

\newcommand{\mypar}[1]{\noindent\textbf{#1}.}

\usepackage{tikz}
\newlength{\leftLength}
\newcommand{\mySummary}[2]{
	\setlength{\leftLength}{0.1cm + (\widthof{#1}/2)}
	{\centering \begin{tikzpicture}
			\node[align=center,draw,thin,minimum width=\linewidth,inner sep=0.9mm] (titlebox)%
			{\parbox{0.95\linewidth}{\vspace*{1.5ex}\noindent{#2}}};\node[fill=white] (W) at ([xshift=\the\leftLength] titlebox.north west) {{\small #1}};%
\end{tikzpicture}}}

\begin{document}

\title{An Experimental Analysis of Graph-Distance Algorithms for Comparing API Usages}

\author{\IEEEauthorblockN{Sebastian Nielebock}
	\IEEEauthorblockA{\ovgu, Germany} 
		\textit{sebastian.nielebock@ovgu.de}
\and
\IEEEauthorblockN{Paul Blockhaus}
	\IEEEauthorblockA{\ovgu, Germany} 
	\textit{paul.blockhaus@ovgu.de}
\and
\IEEEauthorblockN{Jacob Krüger}
	\IEEEauthorblockA{Ruhr-University Bochum,\\ Germany}
	\textit{jacob.krueger@rub.de}
\and
\IEEEauthorblockN{Frank Ortmeier}
	\IEEEauthorblockA{\ovgu, Germany} 
	\textit{frank.ortmeier@ovgu.de}
}

\maketitle

\begin{abstract}
  \looseness=-1
Modern software development heavily relies on the reuse of functionalities through Application Programming Interfaces (APIs). 
However, client developers can have issues identifying the correct usage of a certain API, causing misuses accompanied by software crashes or usability bugs. 
Therefore, researchers have aimed at identifying API misuses automatically by comparing client code usages to correct API usages. 
Some techniques rely on certain API-specific graph-based data structures to improve the abstract representation of API usages. 
Such techniques need to compare graphs, for instance, by computing distance metrics based on the minimal graph edit distance or the largest common subgraphs, whose computations are known to be NP-hard problems. 
Fortunately, there exist many abstractions for simplifying graph distance computation.
However, their applicability for comparing graph representations of API usages has not been analyzed.
In this paper, we provide a comparison of different distance algorithms of API-usage graphs regarding correctness and runtime. 
Particularly, correctness relates to the algorithms' ability to identify similar correct API usages, but also to discriminate similar correct and false usages as well as non-similar usages. 
For this purpose, we systematically identified a set of eight graph-based distance algorithms and applied them on two datasets of real-world API usages and misuses. 
Interestingly, our results suggest that existing distance algorithms are not reliable for comparing API usage graphs. 
To improve on this situation, we identified and discuss the algorithms' issues, based on which we formulate hypotheses to initiate research on overcoming them.

\end{abstract}

\section{Introduction}
\label{sec:introduction}

\looseness=-1
\noindent
Modern software development heavily relies on reusing existing software to effectively and efficiently construct desired products. 
Software reuse can include copying\,\&\,pasting code from other locations or discussion forums (e.g., StackOverflow), (internal) software platforms or product lines, and the integration of specified Application Programming Interfaces (APIs) of (external) software libraries in an ecosystem~\cite{Krueger2020CostComparision,Krueger1992Reuse,Bosch2010ImpactSPLs}. 
Especially the latter is vulnerable to be misused by client developers. 
For instance, a developer may use a method of an API differently than expected by the API developers (e.g., parameters contradicting an implicit specification of that method). 
We call cases in which this causes unexpected negative behavior of the software \emph{API misuses}, which may manifest as crashes, usability problems, or security issues~\cite{Nadi2016, Murali2017, Oliveira2018, Amann2018a}. 
In this paper, we consider API usages as correct or incorrect usages (i.e., misuses), and their comparison as a way to discriminate similar correct API usages from similar misuses as well as to distinguish completely different API usages.

We focus on API misuses since they are a prevalent issue in software development. 
For example, studies show that approximately half of the bug fixes in five open-source projects require an adaptation of API usages~\cite{Zhong2015} and more than half of 806 projects use outdated APIs~\cite{Wang2020} which may cause security issues. 
To describe such API misuses, Amann et al.~\cite{Amann2019} have defined a corresponding taxonomy.
Other studies identified root causes of API misuses, for instance, the absence of proper API documentation, APIs that were too complex, a lack of domain knowledge, backward incompatibilities, issues with the execution environment, or a lack of communication channels between API and client developers~\cite{Robillard2009, Robillard2011, Hou2011, Nadi2016, Nam2019, Patnaik2019, Gorski2020, Lamothe2020, Zhang2020}.

Researchers focus on two directions to deal with API misuses:
First, avoiding them by mitigating the above causes, for instance, using automated tools to enhance the documentation~\cite{Treude2016}.
Second, automatically detecting API misuses.
In this context, techniques for mining and comparing specifications of API usages against the suspicious API client code are prevalent. 
These specifications may be represented as formal specification, such as finite-state automata~\cite{Ammons2002, Gabel2008, Wasylkowski2011} and dynamic invariants~\cite{Ernst2007}, or as patterns of API usages~\cite{Livshits2005, Thummalapenta2007, Nguyen2009a, Allamanis2014, Amann2018a, Nielebock2020}. 

We focus on the second direction, which typically requires comparisons of different API usages. 
This encompasses searching and mining similar API usages~\cite{Saul2007, Moreno2015, Nguyen2017, Chen2019, Nielebock2021a}, comparing pattern candidates~\cite{Nguyen2009a, Amann2018a}, or detecting pattern violations~\cite{Amann2018a, Nielebock2021a, Kang2021}. 
Since API usages are more and more represented as graphs~\cite{Nguyen2009a, Amann2018a, Nielebock2020, Kang2021}, this essentially means to compare the distance of graphs. However, established graph-distance algorithms, such as computing the minimal graph edit distance (GED) or the maximum common subgraph, are NP-hard. 
Fortunately, advances have been made to relax or approximate distance computation. 
Still, to the best of our knowledge, the applicability of these algorithms for comparing API usages has not been systematically analyzed. 

\looseness=-1
In this paper, we address this gap by analyzing a set of well-known graph-distance algorithms and investigating whether they are feasible to compare API usage graphs. 
We consider a \enquote{good} distance algorithm to be \emph{effective} and \emph{efficient}. 
For detecting API misuses, an effective algorithm is able to compute a distance that significantly differs when comparing two correct (or two incorrect) usages rather than comparing a correct usage and a misuse. 
So, we can discriminate correct from incorrect (i.e., misused) API usage, and effectively reduce the false positive rate of misuse detectors, a well-known issue in static API misuse detection~\cite{Goues2012,Amann2019a}. 
For searching similar API usages, the algorithm should compute a low distance value for API usages of the same API in similar contexts. 
This way, the algorithm can help to effectively filter API usages for subsequent pattern mining.
Since pattern mining usually requires multiple thousands to millions of comparisons, the algorithm must efficiently compute the distance of real-world graphs. 
While we do not expect that the comparison is interactively usable (i.e., done in a fraction of seconds), it should be efficient enough to compare several thousands of graphs in a matter of minutes. 
This is comparable to automated tests executed in a continuous integration system~\cite{Ziftci2017}.

For our experimental comparison, we systematically identified eight graph-distance algorithms and selected those able to achieve our goals. 
Then, we selected two different datasets of API misuses and correct usages and transformed each entry into an \emph{API Usage Graph} (AUG), an established data structure proposed by Amann et al.~\cite{Amann2018a, Amann2019a} (cf. \autoref{sec:api-and-graph-similarity}), as well as into so-called API misuse correction rules (cf. \autoref{ssec:cor-rule}) that we have proposed~\cite{Nielebock2020}.  
We computed the distances between all AUGs (i.e., correct to correct usages, misused to misused, misused to correct usages, and vice versa) within a framework we implemented. 
Based on the distribution of the resulting distance values, we determined the ability of the selected algorithms to effectively discriminate misused from correct API usages. 
We discuss our main results as well as their implications in~\autoref{sec:experiment}, and publish our data, results, and experimental framework in our replication package.\footnote{\url{https://doi.org/10.5281/zenodo.5255402}\label{fn:repo}}
\begin{figure}
	\includegraphics[width=\linewidth,trim={1cm 1cm 1cm 1cm},clip]{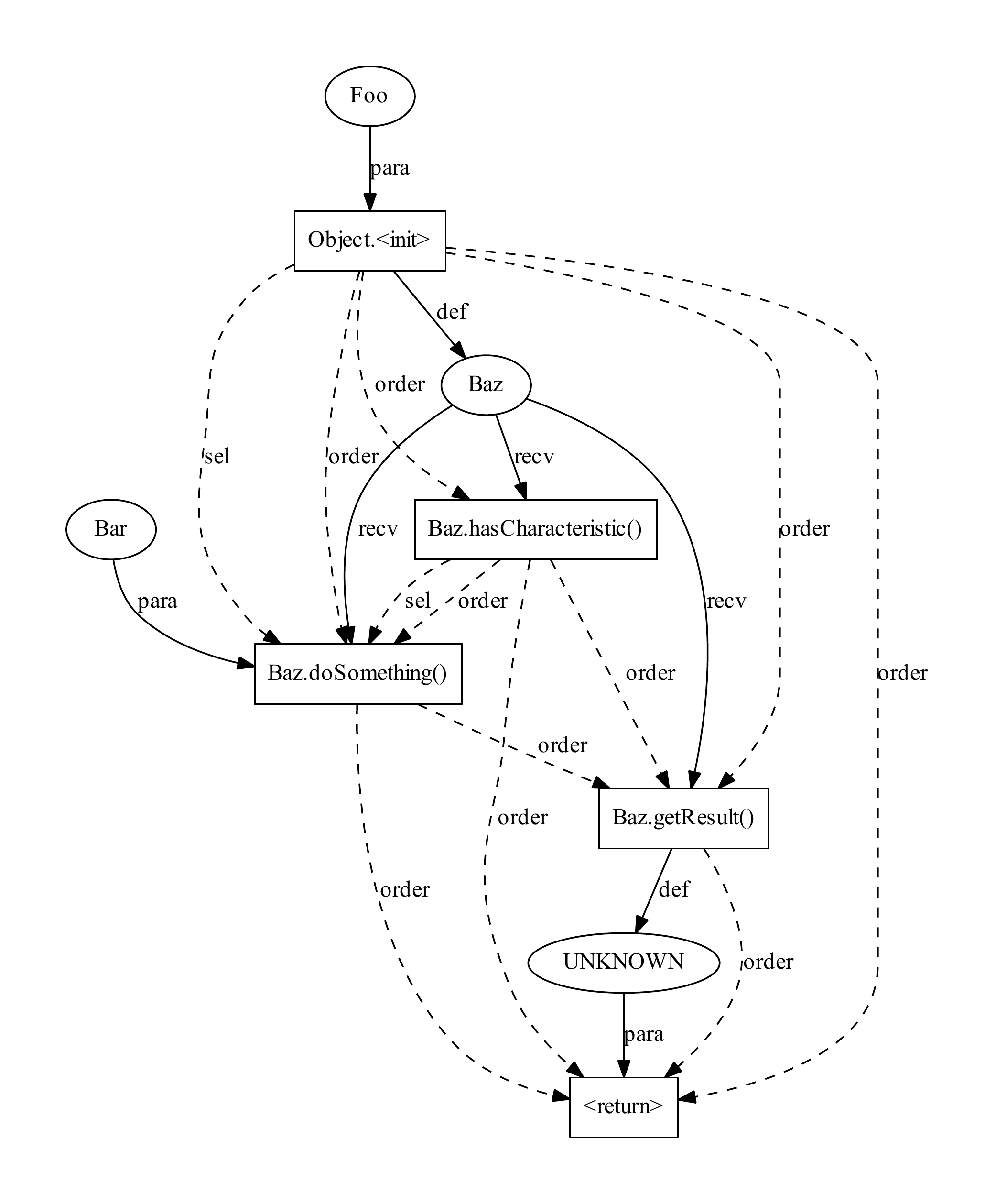}
	\caption{AUG representing the method \texttt{computeSomething} in \autoref{lst:code}.}
	\label{fig:aug-code}
	\vspace*{-2ex}
\end{figure}

\begin{lstlisting}[caption={Code example for the AUG in \autoref{fig:aug-code}.},float,label={lst:code}]
package pkg.at.some.loc;

import from.another.place.Foo;
import from.another.place.Bar;
import from.another.place.Baz;

public class AUGSample {
    Foo fooObj = new Foo(42,"text");

    public Integer computeSomething(Bar barObj) {
        Baz bazObj = new Baz(this.fooObj);
        if(bazObj.hasCharacteristic()){
            bazObj.doSomething(barObj);
        }
        return bazObj.getResult();;
    }
}
\end{lstlisting}

\section{API Usage Representations}
\label{sec:api-and-graph-similarity}

\noindent
In this section, we describe the basic concepts required to understand our experimental comparison.

\subsection{API Usage Graphs (AUG)}
\label{ssec:aug}

\looseness=-1
\noindent
Source code can be represented by different data structures, such as text, token streams, Abstract Syntax Trees (ASTs), control flow graphs, or Program Dependency Graphs (PDGs). 
For API misuse detection, Amann et al.~\cite{Amann2018a, Amann2019a} developed the AUG. 
It is tailored to represent the specificities of API usages, in contrast to, for instance, general-purpose PDGs. 
Using AUGs as pattern representations of API usages, Amann et al. achieved a higher precision and recall than comparable algorithms for detecting API misuses. 
In their recent work, Kang et al.~\cite{Kang2021} defined \emph{extended AUGs (eAUGs)} to further include specific properties of API usages as additional nodes and edges. 
They found that using these extensions together with active learning had positive effects on misuse detection. 
However, for our experimental comparison of graph-distance algorithms, the simpler original AUGs are sufficient since these algorithms do not discriminate between different node and edge types. 
Note that we target the Java programming language with its specific elements, and thus we limit our descriptions to Java, too.
As a running example, we use an AUG (cf. \autoref{fig:aug-code}) based on the method \texttt{computeSomething} starting in line 10 of \autoref{lst:code}. 

In general, an AUG represents a directed, labeled multigraph $aug := (V, E, \Sigma_V,\Sigma_E, s, t, l_V,l_E)$ where $V$ is a set of vertices or nodes, $E: V \times V$ is a multiset of edges, $\Sigma_V$ and $\Sigma_E$ are finite alphabets of node ($V$) and edge ($E$) labels, $s: E \rightarrow V$ and $t: E \rightarrow V$ are functions to map an edge to its source ($s$) or target ($t$) node, and $l_V: V \rightarrow \Sigma_V$ and $l_E: E \rightarrow \Sigma_E$ are the node and edge labeling functions, respectively.
As we can see in \autoref{fig:aug-code}, AUGs may involve multiple types of nodes and edges. 
Nodes can represent actions (i.e., rectangles in the graphical representation) or data (i.e., ellipses in the graphical representation). 
An action node describes an API method call (e.g., node \texttt{Baz.doSomething()}) or a control structure (e.g., node \texttt{<return>}). 
A data node represents a raw value or an object instance (e.g., node \texttt{Baz}). 
There exist several subtypes of these nodes. 
To determine these node types, we define the function $type: V \rightarrow String$, which returns the type of a particular node as a String value. 

An edge can represent either the data-flow (solid arrow) or the control-flow (dashed arrow). 
Data-flow edges can show the usage of a data node as a parameter (e.g., the \texttt{para} edge directing from node \texttt{Bar} to the node \texttt{Baz.doSomething()}), an object instance on which a method is called (e.g., the \texttt{recv} edge directing from node \texttt{Baz} to \texttt{Baz.doSomething()}), or the creation of a new object instance (e.g., the \texttt{def} edge directing from node \texttt{Object.<init>} to node \texttt{Baz}). 
Control-flow edges can show a condition (e.g., the \texttt{sel} edge directing from node \texttt{Baz.hasCharacteristic} to node \texttt{Baz.doSomething}) or certain execution orders (e.g., the \texttt{order} edge between node \texttt{Baz.doSomething} and \texttt{Baz.getResult}). 
For further details on node and edge types, we refer to Sven Amann's dissertation~\cite{Amann2018a}.

\looseness=-1
Since AUGs are generated based on static ASTs, control-flow information and type resolution are limited. 
Particularly, \texttt{order}-edges are generated conservatively, namely as the transitive closure of all \texttt{order}-edges between each pair of action nodes. 
To enable type resolution, the AUG generation requires access to the source code or library (i.e., of the used API) to find the declarative type of, for instance, a certain method. 
If this is not provided, the AUG generation uses the type \texttt{UNKNOWN}. 
Note that the \texttt{UNKNOWN}-node in \autoref{fig:aug-code} cannot be resolved, since the generation did not have access to the declaration of the method \texttt{Baz.getResult()}. 
Thus, it cannot decide whether the object (i.e., the return type of the method \texttt{computeSomething()}) is of type \texttt{Integer} or of a subtype of \texttt{Integer}. Also, the generation fails to correctly resolve dynamically inferred types, such as generic types in Java.

Some graph-distance algorithms rely on node and edge labels to compute similarity, which is why the labeling functions $l_E$ (i.e., edge labeling) and $l_V$ (i.e., node labeling) are important~\cite{Amann2018a}.
Regarding edges, $l_E$ assigns edges their respective types, and thus $|\Sigma_E|$ denotes the number of different edge type names. 
Labels of action nodes describe API method calls, such as \texttt{Baz.doSomething()}. 
This consists of the declaring type name of the called method (i.e., \texttt{Baz}) and the method name (i.e., \texttt{doSomething()}). 
Note that parameters are not part of the label, since they are represented by additional nodes connected with \texttt{para}-edges. 
In case of specialized action nodes, for instance, return statements (i.e., \texttt{<return>}) or object constructors (i.e., \texttt{<init>}), special labels are defined. 
For an overview of these label types, we refer to the original work of Sven Amann~\cite{Amann2018a}. 
Regarding data nodes, the resolved declaring type names are used as labels. 

\looseness=-1
For our analysis, we adapted the standard definition twice. 
First, we used slightly different labels for data nodes. Namely, we label nodes representing raw values of primitive types (e.g., \texttt{int}, \texttt{String}) with the actual value. 
Our rationale is that these values may indicate a certain meaning, which would be hidden when abstracting them with the declaring type name. 
For example, the method \texttt{getInstance} from the class \texttt{java.security.MessageDigest}\footnote{\url{https://docs.oracle.com/en/java/javase/11/docs/api/java.base/java/security/MessageDigest.html}} requires a \texttt{String} representing the hashing algorithm as input.
Second, we define a function $api : V \rightarrow String$ that returns the complete declaring type name if it is resolvable (e.g. \texttt{java.lang.Object} for the node \texttt{Object.<init>}). 
If the type cannot be resolved, the String represents the type name or an empty String in case no type is apparent.

\subsection{AUG Correction Rules}
\label{ssec:cor-rule}
\begin{figure*}
	\centering
	\includegraphics[width=.6\linewidth,trim={1cm 1cm 1cm 1cm},clip]{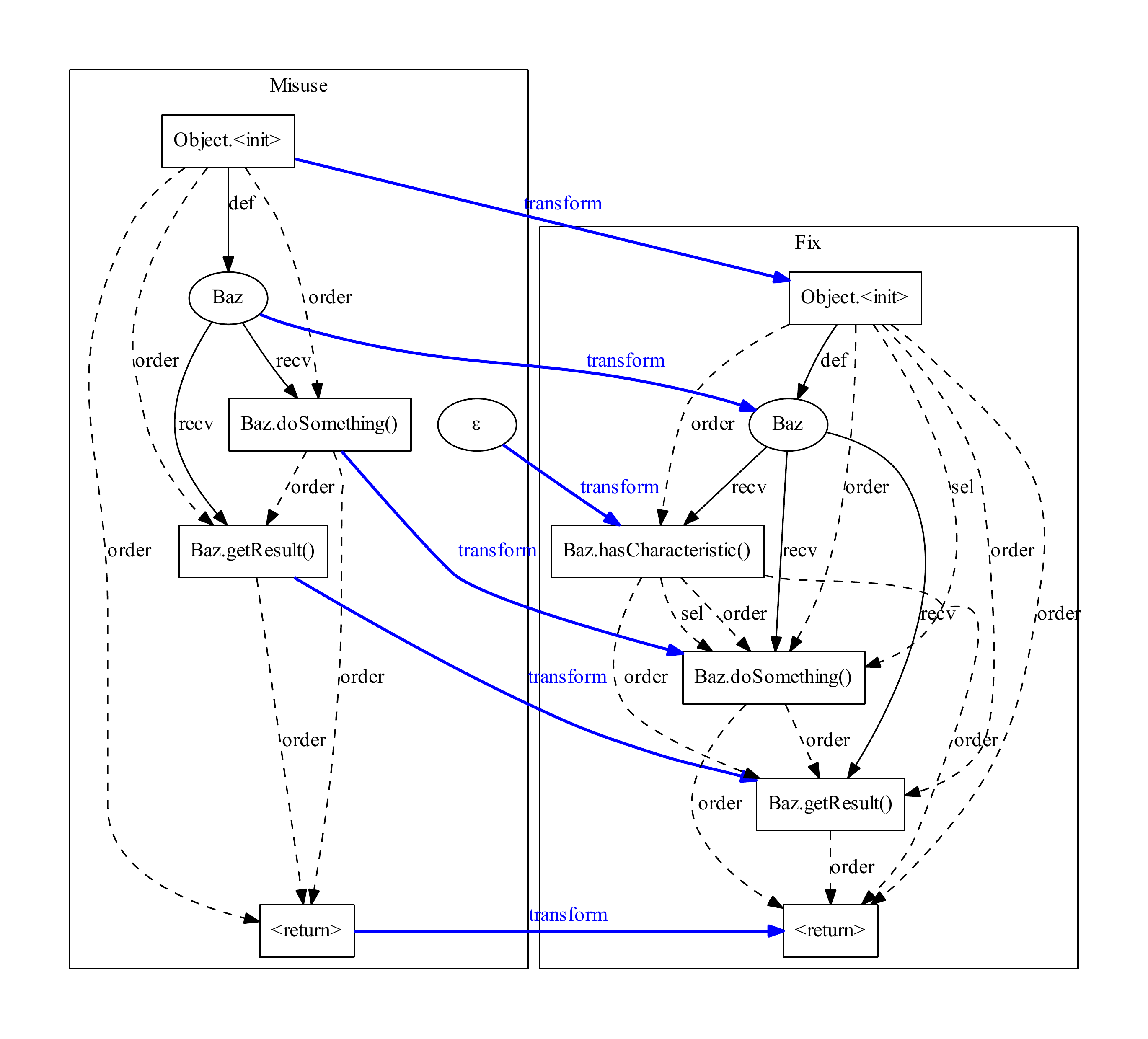}
	\caption{AUG correction rule by adding the condition in line 12 in \autoref{lst:code}.}
	\label{fig:aug-fix-rule}
\end{figure*}

\noindent
In our previous work, we introduced the notion of \emph{correction rules}~\cite{Nielebock2020}: an AUG-based encoding of changes needed to fix an API misuse, which can be automatically generated from fixing commits. 
The goal was to transfer the knowledge of one API-misuse fix to similar usages in other projects. 
We show an example of such a rule in \autoref{fig:aug-fix-rule}. 
This rule describes a case in which a developer forgot to call the condition check in line 12 in \autoref{lst:code} (i.e., \texttt{if(bazObj.hasCharacteristic())}), and defines how to add this call to fix this misuse. 
Each rule consists of two AUGs: a misuse AUG (left) and a fix AUG (right). 
Furthermore, correction rules describe the changes needed to transform the misuse into the fix. 
For that purpose, we computed the minimal mapping (depicted as blue \texttt{transform}-edges) between the two AUGs using the Kuhn-Munkres algorithm~\cite{Munkres1957}. 
To simplify the rule, we left out nodes (with their respective edges) that are not affected by any change. 
For instance, the nodes \texttt{Foo} and \texttt{Bar} in \autoref{fig:aug-code} are not part of the rule in \autoref{fig:aug-fix-rule}. 
Note that for real examples, depending on the committed changes, the number of left out nodes is usually much higher. 
An addition, like the added conditional statement, is represented by mapping a so-called empty node (i.e., \texttt{$\epsilon$}) in the misuse AUG to the respective added node (e.g., \texttt{Baz.hasCharacteristics()}). 
A deletion is represented vice-versa.
From such correction rules, we can derive related misuse and fix AUGs that we can compare to other usages, allowing us to evaluate graph-distance algorithms.
\section{Methodology}
\label{sec:methodology}

\noindent
In the following, we describe the methodology we employed for our experimental analysis.

\subsection{Goal of the Distance Computation}
\label{ssec:goal-dist}

\noindent
We define $dist$ to be a distance function taking as input two AUGs $aug_i$ and $aug_j$ to compute a normalized distance, namely: $dist(aug_i, aug_j) \in [0,1]$. 
$0$ denotes that the two usages described by the AUGs are identical, while $1$ means that the usages are most dissimilar (i.e., completely different API calls). 
We distinguish between two types of API usages, and thus of AUGs: correct usages (i.e., $aug_c$) and misuses (i.e., $aug_m$). 
Moreover, we define two reference usages described by their respective AUGs: a correct one $aug_{rc}$ and a misuse $aug_{rm}$. 
The goal of $dist$ is to compute a distance between a reference AUG and another AUG so that distances between similar types (i.e., correct usages of the same API) have a smaller distance than two dissimilar types (i.e., a correct usage and a misuse). 
Thus, we expect:
\begin{equation}
	\footnotesize
	\label{eq:sim}
	\begin{split}
	dist(aug_{rc},aug_c)<dist(aug_{rc},aug_m)\\
	dist(aug_{rm},aug_c)>dist(aug_{rm},aug_m)
	\end{split}
\end{equation}

Furthermore, we assume that the reference usages are based on a correction rule (cf. \autoref{ssec:cor-rule}). 
We denote this rule as $aug_{rm} \rightarrow aug_{rc}$, where $aug_{rm}$ is the misuse and $aug_{rc}$ is its respective fix. 
To discriminate misuses from correct usages, in addition to \autoref{eq:sim}, $dist$ should satisfy:
\begin{equation}
	\footnotesize
	\label{eq:rule}
	\begin{split}
		dist(aug_{rc},aug_c)<dist(aug_{rm},aug_c)\\
		dist(aug_{rc},aug_m)>dist(aug_{rm},aug_m)
	\end{split}
\end{equation}
So, we can use a correction rule to compute the distance from both, the misuse and the correct usage, to any arbitrary usage. 
If the distance is lower for the misuse than for the correct part, the usage is marked as misuse. 
Ideally, the correction rule could then be used as a patch to fix the misuse.

To assess such rules, we assume a dataset of AUGs describing correct usages $C=\{aug_{c1},aug_{c2},\cdots,aug_{cm}\}$ and a set of misuses $M=\{aug_{m1},aug_{m2},\cdots,aug_{mn}\}$. We denote a rule to be \emph{applicable} with a distance function $dist$, if it satisfies Equations~\ref{eq:rule_1}, \ref{eq:rule_2}, \ref{eq:rule_3}, and \ref{eq:rule_4}:
\begin{equation}
	\footnotesize
	\label{eq:rule_1}
	\begin{split}
		\frac{\sum_{aug_c \in C}{dist(aug_{rc},aug_c)}}{|C|}< \frac{\sum_{aug_m \in M}{dist(aug_{rc},aug_m)}}{|M|}
	\end{split}
\end{equation}
\begin{equation}
	\footnotesize
\label{eq:rule_2}
	\begin{split}
		\frac{\sum_{aug_c \in C}{dist(aug_{rm},aug_c)}}{|C|}> \frac{\sum_{aug_m \in M}{dist(aug_{rm},aug_m)}}{|M|}
	\end{split}
\end{equation}
\begin{equation}
	\footnotesize
\label{eq:rule_3}
	\begin{split}
		\frac{\sum_{aug_c \in C}{dist(aug_{rc},aug_c)}}{|C|}< \frac{\sum_{aug_c \in C}{dist(aug_{rm},aug_c)}}{|C|}
	\end{split}
\end{equation}
\begin{equation}
	\footnotesize
	\begin{split}
		\label{eq:rule_4}
		\frac{\sum_{aug_m \in M}{dist(aug_{rc},aug_m)}}{|M|}> \frac{\sum_{aug_m \in M}{dist(aug_{rm},aug_m)}}{|M|}
	\end{split}
\end{equation}
These equations generalize Equations~\ref{eq:sim} and \ref{eq:rule} by comparing the average distance over a set of other usages. 
More generally, an \emph{applicable} rule produces significantly different distributions of distance values for one set (e.g., $dist(aug_{rc},aug_c)$) than for another set (e.g., $dist(aug_{rc},aug_m)$).

Our goal is to find a distance function $dist$ that maximizes the number of \emph{applicable} rules on a set of generated correction rules based on a given dataset of correct and misuse AUGs.
However, this problem is not trivial, since existing graph-distance metrics usually correlate with general graph similarity. 
This can be distorted if, for example, two AUGs share only a small fraction of a similar usage (e.g., a usage in a different context), and thus still lead to a too large distance value. 
Then, the differences between a misuse and a correct AUG cannot be determined. 
Another issue is that two usages may represent two completely different cases, but the equations may still be randomly satisfied.
Finally, two correct usages may also represent two alternative solutions for the same misuse, so that the distance falsely indicates a misuse (i.e., false positive).

\subsection{Graphs-Distance Algorithms}

\noindent
For our comparison, we considered different graph-distance algorithms that we identified by reviewing surveys on graph similarity~\cite{Bunke2000}, code clone detection~\cite{Koschke2007, Roy2009}, and binary code similarity~\cite{Stauffer2017}.
In addition, we checked a curated list of binary graph-distance algorithms available on GitHub.\footnote{\url{https://github.com/SystemSecurityStorm/Awesome-Binary-Similarity}}
We focused on algorithms that compute a distance between graphs. 
Particularly, we expect the underlying distance metrics to be applicable on AUGs, namely directed labeled multigraphs as defined in Section~\ref{ssec:aug}. 
Moreover, we expect the distance metrics to describe a relative distance (i.e., $dist(aug_1,aug_2) \in [0,1]$) to easily compare different algorithms.
Furthermore, we selected those algorithms that directly and only compute the distance based on the two compared AUGs. 
This way, we aimed to avoid that knowledge from other AUG comparisons is required. 
For instance, a machine-learning algorithm may learn certain features from previous comparisons, helping it to better discriminate other AUGs. 
However, for such an algorithm, we can hardly decide (also compared to other algorithms) whether the results are based on the algorithm itself or the training data. 
Finally, we required an existing implementation that we could apply in our experiment (e.g., via an API) or a sufficiently detailed description of the algorithm to easily re-implement it.
Based on these criteria, we selected the four underlying metrics we introduce in the following.
Note that we used two different versions for the GED and four different versions of the Exas vector algorithms in our experiments (cf. \autoref{sec:experiment}).
For the sake of simplicity, we describe these metrics based on the distance of the two AUGs $aug_A := (V_A, E_A, \Sigma_{V_A},\Sigma_{E_A}, s_A, t_A, l_{V_A},l_{E_A})$ and $aug_B := (V_B, E_B, \Sigma_{V_B},\Sigma_{E_B}, s_B, t_B, l_{V_B},l_{E_B})$.

\mypar{Graph Edit Distance (GED)}
The Graph Edit Distance (GED) describes the minimal costs of edit operations (i.e., replacements, insertions, and deletions of nodes and edges) to transform one graph into another~\cite{Sanfeliu1983}. 
It is widely used to compute inexact matchings of structurally similar graphs, and thus applies for our use case~\cite{Stauffer2017}.
A critical factor to represent a meaningful GED is a properly chosen cost function to determine the mapping between two graphs~\cite{Serratosa2019}.
Assume $i, j \in V_A$ and $k,l \in V_B$ to be nodes of the two AUGs, and $ij \in E_A$ and $kl \in E_B$ to represent their edges.
To compute the GED on AUGs, we selected the following cost functions for node replacement (\autoref{eq:node_replacement_cost}), node deletion and addition (\autoref{eq:node_del_ins_cost}), edge replacement (\autoref{eq:edge_replacement_cost}), as well as edge deletion and addition (\autoref{eq:edge_del_ins_cost}).
\begin{equation} \label{eq:node_replacement_cost}
	\footnotesize
	cost_r(i,k) = \begin{cases}
		0 & \text{if } l_{V_A}(i) = l_{V_B}(k) \land type(i) = type(k)\\
		1 & \text{if } type(i) = type(k) \\
		2 & \text{otherwise}
	\end{cases}
\end{equation}
\begin{equation} \label{eq:node_del_ins_cost}
	\footnotesize
	cost_d(i) = cost_a(k) = 2
\end{equation}
\begin{equation}\label{eq:edge_replacement_cost}
	\footnotesize
	cost_r(ij,kl)_=\begin{cases}
		0 & \text{if } l_{E_A}(ij) = l_{E_B}(kl) \\
		2 & \text{otherwise}
	\end{cases}
\end{equation}
\begin{equation}\label{eq:edge_del_ins_cost}
	\footnotesize
	cost_d(ij) = cost_a(kl) = 2
\end{equation}
Regarding the edge costs (i.e., \autoref{eq:edge_replacement_cost}), we do not need to separate costs for individual types, since the label function $l_E$ already denotes these types (cf. \autoref{ssec:aug}).

Due to the variety of different node and edge types in AUGs, one may define a large number of different cost functions to account for their specific properties. 
For the sake of simplicity, we performed only small modifications to the cost function (i.e., equal costs for all edit operations) and handle all nodes identically. 
Note that our cost functions still satisfy the triangle inequality: $cost_d(i) + cost_a(j) \ge cost_r(i,j)$ and $cost_d(ij) + cost_a(kl) \ge  cost_r(ij,kl)$.
The GED is then defined over all possible sequences of edit operations transforming $aug_A$ into $aug_B$, with the function $ged$ returning the minimal edit costs.
To normalize the distance, we define the maximum costs between nodes (i.e., Equation~\ref{eq:max_node}) and edges (i.e., Equation~\ref{eq:max_edge}).
\begin{equation}
	\footnotesize
	\label{eq:max_node}
	mcost_{n} = \underset{\forall i \in V_A, k \in V_B}{max}\{cost_r(i,k),cost_d(i,k),cost_a(i,k)\}
\end{equation}
\begin{equation}
	\footnotesize
	\begin{split}
	\label{eq:max_edge}
	mcost_{e} =\\\underset{\forall ij \in E_A, kl \in E_B}{max}\{cost_r(ij,kl),cost_d(ij,kl),cost_a(ij,kl)\}
	\end{split}
\end{equation}

For our cost definition, this means $mcost_{n}=mcost_{e}=2$. The normalized distance is then defined as:
\begin{equation}
	\footnotesize
	\begin{split}
	dist_{ged}(aug_A,aug_B) =\\\frac{ged(aug_A,aug_B)}{max(|V_A|,|V_B|) \cdot mcost_{n} + max(|E_A|,|E_B|) \cdot mcost_{e} }
	\end{split}
\end{equation}
The exact computation of $dist_{ged}$ is known to be NP-hard, and thus only applicable for small graphs. 
For this reason, we applied two algorithms using heuristics to compute an almost exact GED. First, the algorithm of Abu-Aisheh et al.~\cite{AbuAisheh2015} uses a simplified version of the well-known A*-algorithm, building on a depth-first search together with a pruning technique to discard edit sequences with high costs. 
Their algorithm is implemented in the NetworkX python library,\footnote{\url{https://networkx.org/documentation/stable/reference/algorithms/generated/networkx.algorithms.similarity.graph_edit_distance.html}} which is why we refer to it as \texttt{NetworkXGED}.

Second, we applied the Hungarian algorithm (also known as Kuhn-Munkres algorithm)~\cite{Munkres1957}. 
This algorithm computes a one-to-one mapping between the nodes of each partition in a bipartite graph, producing minimal edit costs for those nodes. 
Then, this mapping is used to compute the GED. Since this algorithm only considers the costs for node edits, we set $mcost_{e}=0$. 
To compute the minimal mapping, we applied the linear sum assignment implemented in the python library scipy,\footnote{\url{https://docs.scipy.org/doc/scipy/reference/generated/scipy.optimize.linear_sum_assignment.html}} which is based on the description of Crouse~\cite{Crouse2016}. We refer to this algorithm as \texttt{HungarianGED}.

\mypar{Maximum Common Subgraph (MCS)}
Another distance metric is the maximum common subgraph\cite{Bunke1998} (MCS). It is based on the notion that similar graphs share a larger common subgraph. 
A major advantage of this metric is that, in contrast to the GED, it does not require a carefully designed cost function to get a valid and meaningful distance. 
Thus, the MCS qualifies as another candidate to measure the structural similarity of AUGs. Still, it has been shown that the MCS can be calculated with any GED algorithm using the following cost functions for node replacement (\autoref{eq:mcs_node_replace_cost}), node deletion and addition (\autoref{eq:mcs_node_del_cre_cost}), edge replacement (\autoref{eq:mcs_edge_replace_cost}), as well as edge deletion and addition (\autoref{eq:mcs_edge_del_cre_cost})~\cite{Bunke1997}:
\begin{equation}\label{eq:mcs_node_replace_cost}
	\footnotesize
    cost_r(i,k) =\begin{cases}
        0 &  \text{if } l_{V_A}(i) = l_{V_B}(k) \land type(i) = type(k)\\
        \infty & \text{otherwise}
    \end{cases}
\end{equation}
\begin{equation}\label{eq:mcs_node_del_cre_cost}
	\footnotesize
    cost_d(i) = cost_a(k) = 1
\end{equation}
\begin{equation} \label{eq:mcs_edge_replace_cost}
	\footnotesize
    cost_r(ij,kl) =\begin{cases}
        0 & \text{if } l_{E_A}(ij) = l_{E_B}(kl) \\
        \infty & \text{otherwise}
    \end{cases}
\end{equation}
\begin{equation}\label{eq:mcs_edge_del_cre_cost}
	\footnotesize
    cost_d(ij) = cost_a(kl) = 1
\end{equation}

Similar to the $ged$ function, the $mcs$ function computes the minimal costs to produce the maximum common subgraph.
To obtain a normalized distance between $[0,1]$, our $dist_{mcs}$ function is defined as follows:
\begin{equation}\label{eq:dist_mcs}
	\footnotesize
     dist_{mcs}(aug_A,aug_B)  = \frac{mcs(aug_A,aug_B)}{max(|V_A|,|V_B|)+max(|E_A|,|E_B|)}
\end{equation}
Since computing the MCS is also NP-hard, we reuse \texttt{HungarianGED} as \texttt{HungarianMCS} to compute the maximum common subgraph. 
This algorithm includes only node-related costs, and thus we set the denominator of \autoref{eq:dist_mcs} to $max(|V_A|,|V_B|)$ (i.e., ignoring edge costs).

\mypar{Node-Node Similarity}
Another metric for graph similarity is the link-based similarity of nodes originating from hyperlinked environments, such as the world wide web~\cite{Kleinberg1998}. 
However, most algorithms compute similar nodes within one graph only. 
To compare between graphs, Blondel et al.~\cite{Blondel2004} proposed a vertex similarity.
Particularly, they compute a node-node similarity matrix $S$ as the limit of a normalized iterative matrix multiplication with an even number of iterations.  In each iteration, the following formula is applied:
\begin{equation}
	\footnotesize
S_{k+1} = BS_kA^T + B^T S_k A
\end{equation}
$A$ and $B$ are the respective adjacency matrices of $aug_A$ and $aug_B$, and $S_0$ is an all-ones matrix. 
So, the information on the connectivity similarity of the nodes is collected within $S$.

In our framework, we used the existing python implementation in graphsim,\footnote{\url{https://github.com/caesar0301/graphsim}} which is based on NetworkX and call the algorithm \texttt{NodeSimilarityOpt}.
To convert the similarity matrix $S$ into a distance metric, we reduced it with the maximum linear sum assignment ($lsa$) to find the maximum node-node similarities as a set of similarity values. In the second step, we use the average node-node similarity from that result to compute the distance metric as follows:
\begin{equation}
	\footnotesize
    dist_{NodeSim}(aug_A,aug_B)  = 1 -\frac{\sum lsa(S)}{|lsa(S)|)}
\end{equation}

\mypar{Exas-Vectors}
Lastly, we employed distance metrics from the code-clone domain using so-called Exas-Vectors to measure graph distances. 
Nguyen et al.~\cite{Nguyen2009} have shown that Exas-Vectors are able to reasonably approximate the GED, and thus they are eligible to measure the distance of AUGs. 
Exas-Vectors denote vectorizations of graphs whose elements represent the number of certain features present in the graphs. 
Their definition involves two different kinds of features: (p,q)-nodes and n-paths. 
(p,q)-nodes describe the individual nodes (e.g., denoted by their label function $l_V$) together with the number of incoming (i.e., p) and outgoing edges (i.e., q). 
N-paths describe paths of the size $length$ (i.e., number of visited nodes), where nodes and edges are identified via their respective label function (i.e., $l_V$ and $l_E$). 
In our experiments, we omitted n-paths of size one (i.e., single nodes), since they are included in the (p,q)-node feature. 
Moreover, building on the results of Nguyen et al., we limited the maximum path length to four.

To measure the distance between two AUGs, we compute the norm of the difference between their respective Exas-vectors. 
Since the vectors may differ in their specific number and type of features, we first determined the shared features among the two vectors as sub-vectors containing only those features with their respective counts. 
Similarly, we also computed super-vectors containing a union of all features of both vectors, in which the respectively added features are filled up with zeros. 
We then computed two different norms, namely L1-norm and cosine-distance, on the Exas vectors. 
Since the cosine distance is less sensitive to individual differences in the feature count than the L1-norm, we also included the proportion of shared features for this distance. 
We refer to those distances as \texttt{ExasVectorL1Norm} and \texttt{ExasVectorCosine}. 
Assume $vec_A$ and $vec_B$ are the Exas vectors of the AUGs $aug_A$ and $aug_B$, $\hat{vec_A}$ and $\hat{vec_B}$ are the respective super-vectors with all features, $\tilde{vec_A}$ and $\tilde{vec_B}$ the respective sub-vectors with all shared features, $len$ a function to compute the length (i.e., number of elements) of a vector, and $maxVal$ a function to determine the maximum absolute value of a vector. 
Then, we can compute the respective distances as follows:
\begin{equation} \label{eq:exas-l1}
	\footnotesize
	\begin{split}
	dist_{ExasVectorL1Norm}(aug_A,aug_B) =\\ ||\frac{\hat{vec_A}-\hat{vec_B}}{max(1, maxVal(\hat{vec_A}-\hat{vec_B}))} ||_{1}
	\end{split}
\end{equation}
\begin{equation} \label{eq:exas-cos}
	\footnotesize
	\begin{split}
	dist_{ExasVectorCosine}(aug_A,aug_B) =\\ \lambda  \frac{len(\tilde{vec_A})}{len(vec_A)} + (1-\lambda) (1-\frac{\langle\tilde{vec_A}, \tilde{vec_B}\rangle}{||\tilde{vec_A}||_2||\tilde{vec_B}||_2}), 
	\lambda \in [0,1]
	\end{split}
\end{equation}
$\langle\cdot,\cdot\rangle$ denotes the scalar product of two vectors. In our experiments we set $\lambda=0.5$.

We also constructed two additional distance algorithms by splitting the AUGs into subgraphs. 
For this purpose, we cluster nodes of an AUG based on their related packages determined by the $api$ function (cf. Section~\ref{ssec:aug}). 
Then, we construct each subgraph by removing all nodes (together with their connected edges) from the original AUG that does not belong to that cluster. 
So, we obtain a list of subgraphs from a single AUG, each related to a certain API package and, regarding nodes for which no package could be determined, a special miscellaneous subgraph. 
Finally, we compute the distance between two AUGs by computing the distances of their subgraphs that belong to the same API package and averaging all resulting distances.
During this mean computation, we ignore subgraphs that do not have a counterpart in the other AUG (i.e., those that depict a different API usage) as well as sub-distances that equal one. 
The rationale is that these values typically distort the distance computation with noise introduced by unrelated API usages.
We reused the L1-norm and the cosine distance from above and refer to those sub-graph distance algorithms as \texttt{ExasVectorSplitL1Norm} and \texttt{ExasVectorSplitCosine}, respectively.
\section{Experiment}
\label{sec:experiment}

\noindent
Next, we describe our experiments and discuss their results.

\subsection{Data and Experimental Setup}

\noindent
We analyzed the described distance algorithms based on two datasets. 
First, we used \emph{MUBench}\footnote{\url{https://github.com/stg-tud/MUBench}} by Amann et al.~\cite{Amann2016}.
This dataset contains a set of fixed API misuses together with their repository information, the fixing commit, the fixed method, and more details. 
We selected all 116 misuse entries that are linked to a git repository, provide the fixing commit hash, as well as the containing method and source file path. 
Based on these entries, we generated AUGs of the respective misuse (i.e., commit before the fix) and the correct usage (i.e., commit after the fix) as well as the respective correction rule as described in our previous work~\cite{Nielebock2020}. 
Instead of manually determining the import statements to generate rules, we automatically included all external import statements, namely those not starting with the same prefix as the package of the analyzed source file. 
We could generate AUGs and corresponding correction rules for 96 of the 116 misuses (e.g., some source files could not be properly parsed).

The second dataset, \emph{AU500},\footnote{\url{https://github.com/ALP-active-miner/ALP}} has been published by Kang et al.~\cite{Kang2021} and includes manually labeled correct API usages and API misuses. 
This dataset was constructed to provide an independent baseline to assess API-misuse detection tools. 
It comprises 500 API usages, 385 of which are correct usages, while the other 115 represent misuses. 
We were able to generate AUGs for 493 entries (114 misuses and 379 correct usages).

We transformed all generated AUGs and their correction rules into a dot-representation,\footnote{\url{https://graphviz.org/doc/info/lang.html}} including their respective information on the API as well as the types and labels of nodes and edges. 
This way, we could conduct all of our experiments in the same python-based framework by converting and processing AUGs and correction rules with the NetworkX library.\footnote{\url{https://networkx.org/}}
Within our framework, we analyzed the \emph{effectiveness} and the  \emph{efficiency} of the distance algorithms. 
For both, we first generated the correction rules from the MUBench dataset as reference usage in the form $aug_{rm} \rightarrow aug_{rc}$, where $aug_{rm}$ represents the misuse part of the misuse and $aug_{rc}$ its respective fix. 
Moreover, we denote the set of misuse AUGs as $M_{MUBench}$ and the set of correct usages as $C_{MUBench}$. 
For each rule, we then computed the analyzed distance function $dist_x$ to assess the rule's \emph{applicability} as discussed in \autoref{ssec:goal-dist}. 
To this end, we computed the four individual distances $dist_x(aug_{rc},aug_{c})$, $dist_x(aug_{rc},aug_{m})$, $dist_x(aug_{rm},aug_{c})$, and $dist_x(aug_{rm},aug_{m})$, where $aug_{c} \in C_{MUBench}$ and $aug_{m} \in M_{MUBench}$. 
When computing these four distance values, we measured the time using python's \texttt{time} library.\footnote{\url{https://docs.python.org/3/library/time.html}} 
For the algorithm \texttt{NetworkXGED}, we defined a timeout for individual computations, which we set to 15 seconds. 
We selected this value since it ensures that the maximum time to compute all four distances would be at one minute, which was comparable to the execution time of the other algorithms.

Regarding \emph{effectiveness}, we checked for all computed metrics whether each analyzed rule is \emph{applicable} (i.e., satisfies Equations~\ref{eq:rule_1}, \ref{eq:rule_2}, \ref{eq:rule_3}, and \ref{eq:rule_4}). 
We then counted the number of all satisfying rules per distance algorithm (i.e., on MUBench). 
To further assess a rule's ability to detect and fix misuses, we computed the distance values to AU500.
More detailed, we checked each entry of AU500 for which the condition in \autoref{eq:rule} holds, and thereby decided if this entry was identified as a misuse. 
For a reference rule $aug_{rm} \rightarrow aug_{rc}$, this means that an entry AUG $aug_e$ in AU500 satisfies the condition $dist_x(aug_{rc},aug_{e}) > dist_x(aug_{rm},aug_{e})$. 
Since the entries in AU500 are manually labeled as misuse or correct usage, we can compare this result against the labeled ground-truth and compute the rule's precision and recall.
To assess the \emph{efficiency}, we measured the execution time when computing the four distances for each entry on the MUBench dataset.
We provide our experimental data, the framework, and all other scripts in our replication package.\textsuperscript{\ref{fn:repo}}

\subsection{Effectiveness}
\label{ssec:res_effect}

\begin{table}
	\centering
	\caption{Number of \emph{Applicable} rules}
	\label{tab:sat_rules}
	\begin{tabular}{lrcl}
		\toprule
		\textbf{distance} & \textbf{\#rules-sat} & \textbf{/} & \textbf{\#rules} \\
		\midrule
		ExasVectorL1Norm &              0&/&96 \\
		ExasVectorCosine &              6&/&96 \\
		ExasVectorSplitL1Norm &              2&/&96 \\
		ExasVectorSplitCosine &              1&/&96 \\
		HungarianGED &              0&/&96 \\
		NetworkXGED &             15&/&96 \\
		HungarianMCS &              0&/&96 \\
		NodeSimilarityOpt &              0&/&92 \\
		\bottomrule
	\end{tabular}
\end{table}

\begin{table*}
	\centering
	\caption{Precision and Recall of applicable rules together with their applied distance metric and the number of true and false positives/negatives (i.e., \#tp, \#fp, \#tn, \#fn)}
	\label{tab:au500}
	\begin{tabular}{llrrrrrr}
		\toprule
		\textbf{distance} & \textbf{rule\_id} & \textbf{\#fp} & \textbf{\#tp} & \textbf{\#fn} & \textbf{\#tn} &  \textbf{precision} & \textbf{recall} \\
		\midrule
		ExasVectorCosine &               1\_TuCanMobile &   0 &   0 & 114 & 379 &      0.0\% &   0.0\% \\
		ExasVectorCosine &             2\_alibaba\_druid &  26 &   5 & 109 & 353 &    16.13\% &  4.39\% \\
		ExasVectorCosine &                 30\_visualee &  74 &  17 &  97 & 305 &    18.68\% & 14.91\% \\
		ExasVectorCosine &          390\_paho.mqtt.java &  28 &   6 & 108 & 351 &    17.65\% &  5.26\% \\
		ExasVectorCosine &                    473\_ntru &   3 &   1 & 113 & 376 &     25.0\% &  0.88\% \\
		ExasVectorCosine &                   56\_2\_gora &   2 &   3 & 111 & 377 &     60.0\% &  2.63\% \\
		ExasVectorSplitL1Norm &       1\_Apache\_Commons\_Math &  42 &   5 & 109 & 337 &    10.64\% &  4.39\% \\
		ExasVectorSplitL1Norm &             1\_Mozilla\_Rhino &   0 &   0 & 114 & 379 &      0.0\% &   0.0\% \\
		ExasVectorSplitCosine &             1\_Mozilla\_Rhino &   0 &   0 & 114 & 379 &      0.0\% &   0.0\% \\
		NetworkXGED &       1\_Apache\_Commons\_Lang &  52 &  14 & 100 & 327 &    21.21\% & 12.28\% \\
		NetworkXGED &       1\_Apache\_Commons\_Math &  39 &  10 & 104 & 340 &    20.41\% &  8.77\% \\
		NetworkXGED &          1\_Closure\_Compiler &  16 &   3 & 111 & 363 &    15.79\% &  2.63\% \\
		NetworkXGED & 1\_Onosendai\_-\_A\_Better\_Deck &   8 &   2 & 112 & 371 &     20.0\% &  1.75\% \\
		NetworkXGED &      1\_Screen\_Notifications &   3 &   4 & 110 & 376 &    57.14\% &  3.51\% \\
		NetworkXGED &     1\_WordPress\_for\_Android &  94 &  20 &  94 & 285 &    17.54\% & 17.54\% \\
		NetworkXGED &                 29\_visualee &  41 &   9 & 105 & 338 &     18.0\% &  7.89\% \\
		NetworkXGED &       2\_Apache\_Commons\_Lang &  47 &   9 & 105 & 332 &    16.07\% &  7.89\% \\
		NetworkXGED &       2\_Apache\_Commons\_Math &  24 &   5 & 109 & 355 &    17.24\% &  4.39\% \\
		NetworkXGED &          2\_Closure\_Compiler &  24 &   6 & 108 & 355 &     20.0\% &  5.26\% \\
		NetworkXGED &             2\_alibaba\_druid &  31 &  11 & 103 & 348 &    26.19\% &  9.65\% \\
		NetworkXGED &                 30\_visualee &  61 &  11 & 103 & 318 &    15.28\% &  9.65\% \\
		NetworkXGED &               361\_Joda-Time &  22 &   5 & 109 & 357 &    18.52\% &  4.39\% \\
		NetworkXGED &   39\_gae-java-mini-profiler &  72 &  19 &  95 & 307 &    20.88\% & 16.67\% \\
		NetworkXGED &          3\_Closure\_Compiler &  23 &   5 & 109 & 356 &    17.86\% &  4.39\% \\
		\bottomrule
	\end{tabular}
\end{table*}

\noindent
In \autoref{tab:sat_rules}, we summarize the number of \emph{applicable} rules per distance algorithm. 
We can see that only the Exas vector algorithms and \texttt{NetworkXGED} algorithm found \emph{applicable} rules. 
However, even in the best case (i.e., \texttt{NetworkXGED}), they found only a minority of 15 out of 96 possible rules. 
Note that for the \texttt{NodeSimilarityOpt}-algorithm, we could not compute the distance for four rules, and thus the number of checked rules is lower.
Overall, we could identify 24 rules (20 unique ones) for four different metrics.

We then checked these 24 rules against the AU500 dataset using the respective algorithm to compute the distance. 
Based on the number of true and false positives, we computed the precision and recall, which we depict in \autoref{tab:au500}. 
Overall, we can see that for rules that detected at least one misuse (i.e., with $\#tp>0$ or $\#fp>0$), the precision and recall are constantly very low. 
In comparison to the results obtained by Kang et al.~\cite{Kang2021}, who applied MUDetect (precision $27.6\%$, recall $29.6\%$) and ALP (precision $44.7\%$, recall $54.8\%$) using the same dataset, the simple distance metrics could not successfully discriminate misuses from correct usages. 
While we expected our rules to achieve a low recall, since they describe very specific fixes, they could not achieve the aspired high precision.

\subsection{Efficiency}
\label{ssec:res_effic}
\begin{table}
	\centering
	\caption{Mean/median times of the efficiency results on MUBench}
	\label{tab:time}
	\begin{tabular}{lrr}
		\toprule
		\textbf{distance} &  \textbf{time in sec (mean)} &  \textbf{time in sec (median)} \\
		\midrule
		ExasVectorCosine &            1.363347 &              0.046309 \\
		ExasVectorL1Norm &            1.442757 &              0.049212 \\
		ExasVectorSplitCosine &            1.163106 &              0.061127 \\
		ExasVectorSplitL1Norm &            1.177104 &              0.062430 \\
		HungarianGED &            0.010174 &              0.006016 \\
		HungarianMCS &            0.007411 &              0.003391 \\
		NetworkXGED &           15.552184 &              0.323910 \\
		NodeSimilarityOpt &            0.043162 &              0.034601 \\
		\bottomrule
	\end{tabular}
\end{table}

\begin{figure}
	\includegraphics[trim={0.5cm 0 3.5cm 3cm},clip, width=\linewidth]{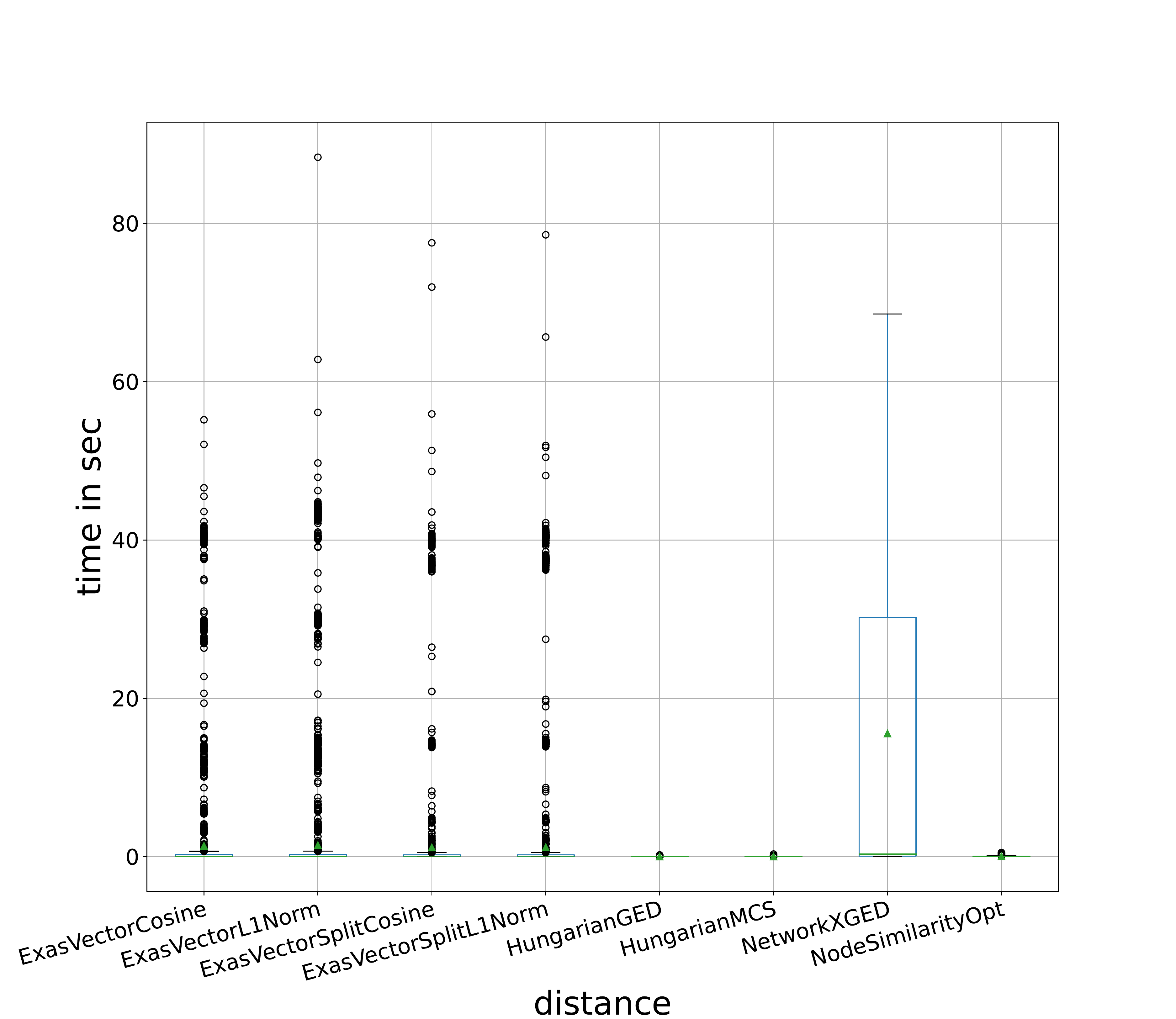}
	\caption{Boxplot of the efficiency results on MUBench}
	\label{fig:time_box}
\end{figure}

\noindent
Based on our time measurements, we obtained time distributions (cf. \autoref{fig:time_box}) as well as median and mean times (cf. \autoref{tab:time}) to compute the four mentioned distance values per rule. 
Based on our results, we observe that most algorithms require roughly one second except for \texttt{NetworkXGED}, which has a mean time of $\approx$ 15 seconds. 
This mean value is heavily influenced by a smaller number of long-lasting computations. Particularly, 1,841 of 5,467 distance computations last longer than 2 seconds.
So, most algorithms can efficiently compute a distance except for \texttt{NetworkXGED}. 
Unfortunately, this algorithm was the one that found most \emph{applicable} rules in our dataset (cf. Section~\ref{ssec:res_effect}). 

\subsection{Root Causes of Low Precision and Recall}

\noindent
Our results indicate that none of the algorithms we selected is sufficient (regarding precision and recall) to effectively discriminate correct usages from misuses.
Thus, we conducted a qualitative, in-depth analysis of the root causes for the low precision and recall. 
In detail, for each algorithm, two reviewers analyzed a sample set of ten pairs of a correction rule $aug_{rm} \rightarrow aug_{rc}$ together with the complete misuse AUG $aug_{m}$ and its respective corrected AUG $aug_{c}$ originating from another misuse-correction pair (all from the MUBench dataset). 
For those samples, the conditions formulated in Equations~\ref{eq:sim} and \ref{eq:rule} must hold. 
Moreover, we ensured that $rm \neq m$ and $rc \neq c$ to avoid that a correction rule is compared to itself.
We were able to analyze such pairs for all metrics, but \texttt{HungarianGED} and \texttt{HungarianMCS}. 
For those algorithms, we did not obtain any entry satisfying our previously mentioned conditions. 
Considering \texttt{NetworkXGED}, nine out of 10 entries have a correction rule and a misuse/correction AUG stemming from the same project. 
Therefore, we re-sampled this set and allowed only entries from different projects. 

The two reviewers individually analyzed the samples to decide whether the usages contain at least one identical API-method call (agreement in 60 out of 70 cases), identify whether this method call is used in a similar context (agreement in 58 out of 70 cases), and add a textual discussion justifying their decision. 
For instance, assume a misuse of an unlabeled button, which was fixed by a correction rule adding a textual label. 
Further, a similar correction fixed the same issue by adding an image to the empty button. 
Then, the reviewer comparing these entries would decide that the same API method call is used, for example, the constructor of the button, and the correction is done in the same context, namely labeling a button.
Finally, the reviewers discussed their decisions with each other and identified the root causes of falsely matched entries.
Next, we discuss general and metric-specific causes for false misuse detection.
For each cause, we formulate hypotheses, which have to be evaluated in subsequent research.

\mypar{General} 
We observed that almost all metrics tend to work better when comparing API usages in the same project than with an external usage. 
This seems to be reasonable, since an API usage may share more structural similarities to a usage in the same project than to an external usage. 
Thus, \emph{applicable} rules may hardly apply to external projects: 

\mySummary{Hypothesis 1}{API misuse detection with correction rules using distance metrics will be more precise if it is applied within the same project than on an external one.}

\mypar{GED} 
We obtained most results for the \texttt{NetworkXGED} algorithm. 
This algorithm had issues with the large number of \texttt{order} edges in AUGs. 
Particularly, many \texttt{order}-edges are \enquote{recycled} without any further costs in the GED computation. 
Therefore, the differences in the node labels, which we perceived as more important for detecting API misuses, are under-represented. For this reason, we argue that:

\mySummary{Hypothesis 2}{API misuse detection with correction rules using the GED is more precise if the costs are adapted to edits of certain AUG element types.}

\mypar{MCS} 
We obtained no rules satisfying the sampling condition, and thus we conclude that computing graph distances using MCSs may not be a valid metric for detecting API misuses.

\mypar{Node-Node Similarity} 
In our analysis, we noticed that this algorithm usually computes a low distance (i.e., high similarity) in case the graphs share nodes that are similarly connected to each other; even though the respective node labels differ. 
For instance, we observed many matches that were caused by handling an exception even though the exception and its causing API differ. 
This happens since nodes, which handle an exception, tend to share structural similarities (e.g., catch-blocks, initialization of an exception object).
Based on this insight, we hypothesize:

\mySummary{Hypothesis 3}{API misuse detection with correction rules using the Node-Node Similarity is more precise when including the similarity of the node labels in the computation.}

\mypar{Exas Vectors} We determined two issues regarding the Exas-vector algorithms. 
First, in some cases, the similarity originates from trivial and individual features that match, such as (p,q)-nodes of \texttt{<return>} or \texttt{<throw>} nodes. 
This matches many non-similar API usages (e.g., since many API usages contain \texttt{<return>}-nodes), which is why:

\mySummary{Hypothesis 4}{API misuse detection with correction rules using the Exas Vectors is more precise when including only features from the vector containing relevant API information.}

\looseness=-1
Second, especially for the \texttt{ExasVectorL1Norm} algorithm, we found many cases in which a high frequency of a single feature diminishes the effect of other differences of the two vectors. 
Particularly, assume two super-vectors $v_1=(1,0,1)$ and $v_2=(0,1,0)$ from their respective Exas vectors in which the position in the vector denotes a unique feature. 
Then, $dist_{ExasVectorL1Norm}(v_1,v_2) = 1$ (i.e., maximum distance), which is reasonable since both vectors share no feature. 
However, assume $v_2'=(0,2,0)$, then $dist_{ExasVectorL1Norm}(v_1,v_2') = \frac{2}{3}$. 
In fact, the normalization of the subtraction $v_1-v_2'$ causes a decreasing distance value even though both vectors still do not share any feature. 
To avoid normalization, one may substitute the frequencies with simple indicators (i.e., 0 if the feature is absent and else 1), so:

\mySummary{Hypothesis 5}{API misuse detection with correction rules using the  \texttt{ExasVectorL1Norm} metric is more precise when using Exas vectors with indicators rather than frequencies.}

Finally, we investigated the effect of splitting AUGs into API-specific sub-graphs. 
We found that splitting usually improves the results, while the corresponding metrics still suffer from the previously mentioned problems.
As a consequence, we argue:

\mySummary{Hypothesis 6}{API misuse detection with correction rules using the Exas vectors is more precise when splitting graphs into API-specific subgraphs.}
\section{Threats to Validity}
\label{sec:threats}

\noindent
Following the classification by Siegmund et al.~\cite{Siegmund2015}, we consider threats to the \emph{internal} and \emph{external} validity. 

\subsection{Internal Validity}

\noindent
Threats to the internal validity describe phenomena of our setup that may harm the trustfulness of the results. 
First, our implementation may contain errors that eventually lead to null results. 
While we cross-checked and tested our implementation on sample data, we still cannot ensure its correctness. Moreover, conceptional issues with the algorithms can taint the results (e.g., different methods produced by overloading could result in the same features in Exas vectors).
Thus, for transparency and replication, we publish our code.\textsuperscript{\ref{fn:repo}}
Second, whether a misuse in the original dataset represents a real misuse is dependent on the original decision of the respective authors. 
Since we did not re-check the validity of the ground truth, this may influence the results.
Third, even though we carefully reviewed existing literature, we have arguably not identified, nor used, all existing algorithms---which may perform better than the analyzed ones. 
However, this could be easily fixed by integrating and comparing other algorithms in our published framework.
Finally, even though the reviewers individually performed the root-cause analysis and discussed their decisions with each other, it is still a subjective view. 
Other reviewers may find different issues or argue that our derived hypotheses are not reasonable.
Thus, we recommend further experiments to validate our hypotheses, and potentially derive new ones.

\subsection{External Validity}

\noindent
External validity considers phenomena that may prevent us from generalizing our results.
In our experiments, we restricted the applicability of the algorithms on API usage graphs, and thus on the Java programming language. 
While we expect that AUGs can be adapted to other programming languages, we cannot ensure that the results apply to other languages. 
Moreover, we cannot state whether the results apply to other graph types, such as control-flow graphs or the extended AUGs introduced by Kang et al.~\cite{Kang2021}. 
This will be subject of further research.
Finally, the MUBench dataset may not be a representative set of API misuses, which could cause their limited applicability to detect misuses in the AU500 dataset. 
Thus, other datasets should be researched, such as the ones provided by Kechagia et al.~\cite{Kechagia2021} or ourselves~\cite{Nielebock2021}.
\section{Related Work}
\label{sec:related-work}

\noindent
Our work relates to surveys on code similarity techniques, graph similarity, and comparative studies of automated software-engineering techniques.

\subsection{Surveys on Code Similarity}

\noindent
Code similarity is actively used for code-clone detection~\cite{Koschke2007,Roy2009}, analyzing plagiarism and software license compliance~\cite{Ragkhitwetsagul2018a}, code search~\cite{Holmes2005, Sahavechaphan2006, Asyrofi2020}, code quality analysis~\cite{Ragkhitwetsagul2018a}, as well as automated program repair (e.g., plastic surgery hypothesis)~\cite{LeGoues2019}.
Most prominently, Haq et al.~\cite{Haq2021} surveyed 70 techniques on binary code similarity regarding their applications, characteristics, implementation, benchmarks, and evaluation techniques. 
They focus on techniques for binary code, and thus deal with additional noise, for instance, by different compiler settings or obfuscation. 
Also, Haq et al. did not perform comparative experiments.
Ragkhitwetsagul et al.~\cite{Ragkhitwetsagul2018a} analyzed 30 different code similarity techniques regarding their applicability to certain degrees of code changes, for instance, global and local changes. 
The study focuses on detecting copied and modified code, while our study refers to detecting similarities between correct and false API usages. 
Moreover, our study involves AUGs as intermediate code representation, while Ragkhitwetsagul et al. used text similarity and simpler code structures, such as ASTs.

\subsection{Surveys on Graph Similarity}

\noindent
While driven by the respective domains in which graph-similarity algorithms are needed, some general surveys on graph similarity exist. 
For instance, Aggarwal~\cite{Aggarwal2015} discusses the general problem of (sub-)graph isomorphism and graph distance together with state-of-the-art algorithms in his data mining book (cf. Chapter 17). 
Gao et al.~\cite{Gao2010} provide a survey on algorithms computing the graph edit distance. 
We used these works as sources to systematically select candidate distance algorithms. 
Ren et al.~\cite{Ren2018} observed that the performance of subgraph isomorphism algorithms is correlated to the structure of graphs. 
Thus, we directly analyzed the performance of realistic AUGs and the respective AUG correction rules.
Other surveys consider more advanced metrics, such as parallel algorithms to compute graph distances~\cite{Kollias2014} or deep learning techniques~\cite{Ma2021}. 
We did not consider these algorithms, since they require more complex computation and hardware as well as carefully selected training sets. 
However, in case our hypotheses are shown to not be true, will not significantly improve the results, or suffer from poor performance, we will analyze more complex algorithms.

\subsection{Studies on Automated Software-Engineering Techniques}

\looseness=-1
\noindent
Our work aligns with different comparative studies on techniques in the automated software-engineering domain, such as automated program repair in general~\cite{Durieux2019}, automated repair of API misuses~\cite{Kechagia2021}, static code analysis for detecting security vulnerabilities~\cite{GosevaPopstojanova2015}, fault localization techniques~\cite{Zou2021}, or the performance of API misuse detectors~\cite{Amann2019}.
To the best of our knowledge, no previous work exists that evaluates graph-distance algorithms for API usages comparisons.
So, even though our results did not indicate a clear benefit of such algorithms, we still contribute to the existing body of knowledge.

\section{Conclusion}
\label{sec:conclusion}

\noindent
Using APIs is accompanied by their potential misuses causing negative effects in the implemented software. 
Thus, researchers developed automated techniques to detect such misuses. 
For this purpose, recent techniques represent API usages as AUGs and compare usages to known correct usages or misuses to detect and ideally repair misuses. 
However, to the best of our knowledge, no studies analyzing existing graph-distance algorithms to detect API misuses exist. 
Therefore, in this paper, we formalized this problem and the desired goal, introduced a set of well-known algorithms, and conducted a study using two independent data sets of real API misuses. 
Our results indicate that the analyzed algorithms fail to effectively discriminate correct API usages from misuses. 
Based on a post-analysis, we identified the potential issues of individual algorithms and derived six hypotheses for further improvements. 
In the future, we will analyze these hypotheses and investigate more advanced techniques for graph distance computation.

\balance
\bibliographystyle{IEEEtranS}
\bibliography{publishers,fullBib,MYfull,paper-api-usage-similarity} 
\end{document}